\documentclass[11pt]{article}

\usepackage[final]{acl}

\usepackage{times}
\usepackage{latexsym}

\usepackage[T1]{fontenc}

\usepackage[utf8]{inputenc}

\usepackage{microtype}

\usepackage{inconsolata}

\usepackage{graphicx}

\usepackage{amsmath,amssymb}
\usepackage{booktabs}
\usepackage{tikz}
\usetikzlibrary{arrows.meta,positioning}

\newcommand{\kap}{\kappa}
\newcommand{\eg}{e.g.,\ }


\title{Bayesian Belief Layer for Controllable Opinion Dynamics in LLM Agents}

\author{
  \textbf{Hafsa Akbar\textsuperscript{1}},
  \textbf{Daniel Platnick\textsuperscript{2,3}},
  \textbf{Marjan Alirezaie\textsuperscript{2,3}}, \\
  \textbf{Hossein Rahnama\textsuperscript{1,2,3}},
  \textbf{Alex "Sandy" Pentland\textsuperscript{1}}
\\
  \textsuperscript{1}MIT Media Lab, Massachusetts Institute of Technology
\\
  \textsuperscript{2}Flybits Labs, Creative AI Hub
\\
  \textsuperscript{3}Toronto Metropolitan University
\\
 \small{\textbf{Correspondence:} \href{mailto:hafsa@mit.edu}{hafsa@mit.edu}}
}

\begin{document}
\maketitle
\begin{abstract}
LLM agents in social simulation revise their opinions implicitly, in context: how open an agent is to persuasion can neither be specified nor verified, and collective outcomes inherit the model's training prior. We introduce Bayesian Chronicle Agents (BCA), a minimal belief layer separating \emph{what} an agent believes from \emph{how} it speaks. Each stance is a probability, updated by one Bayesian step per utterance heard. A single prior-strength parameter $\kap$ encodes stubbornness, modeled after its role in Friedkin--Johnsen (FJ) opinion dynamics. We then sweep this parameter to yield three canonical regimes of opinion dynamics on demand (consensus, persistent disagreement, committed-minority influence), with persistent disagreement matching the FJ closed-form fixed points at $R^2\!=\!0.93$--$0.99$. We further show that prescribed $\kap$ remains recoverable after the language round-trip, with perfect rank-order recovery across all four models. Explicit belief also makes simulation auditable: the layer surfaces systematic per-model stance biases that end-to-end simulation would silently absorb.
\end{abstract}

\section{Introduction}
\label{sec:intro}
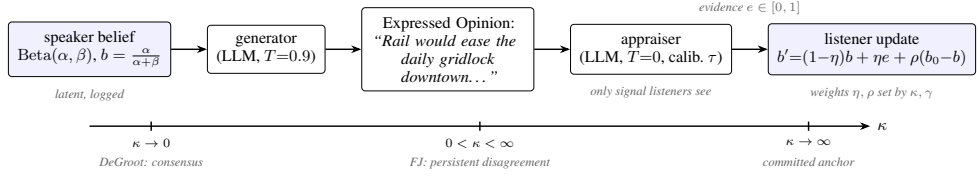
\begin{figure*}[t]
\centering
\resizebox{0.8\textwidth}{!}{%
\begin{tikzpicture}[
  font=\small,
  box/.style={draw, rounded corners=2pt, align=center, inner sep=5pt,
              minimum height=9mm},
  lab/.style={font=\scriptsize\itshape, text=black!60},
  arr/.style={-{Stealth[length=2mm]}, thick}
]
\node[box, fill=blue!6] (bel) {speaker belief\\
  $\mathrm{Beta}(\alpha,\beta)$, $b=\tfrac{\alpha}{\alpha+\beta}$};
\node[box, right=7mm of bel] (gen) {generator\\(LLM, $T{=}0.9$)};
\node[box, right=7mm of gen, text width=30mm]
  (utt) {Expressed Opinion: \itshape``Rail would ease the\\daily gridlock downtown\ldots''};
\node[box, right=7mm of utt] (app) {appraiser\\(LLM, $T{=}0$, calib.\ $\tau$)};
\node[box, right=7mm of app, fill=blue!6] (upd)
  {listener update\\$b' {=} (1{-}\eta)b + \eta e + \rho(b_0{-}b)$};
\draw[arr] (bel) -- (gen);
\draw[arr] (gen) -- (utt);
\draw[arr] (utt) -- (app);
\draw[arr] (app) -- node[above=5.5mm, lab] {evidence $e\in[0,1]$} (upd);
\node[lab, below=1mm of bel] {latent, logged};
\node[lab, below=1mm of app] {only signal listeners see};
\node[lab, below=1mm of upd]
  {weights $\eta,\rho$ set by $\kap,\gamma$};
\begin{scope}[yshift=-15mm]
\draw[thick,-{Stealth[length=2mm]}] (0,0) -- (15.2,0)
  node[right, font=\small] {$\kap$};
\foreach \x/\l/\d in {1.2/{$\kap\to 0$}/{DeGroot: consensus},
                      7.6/{$0<\kap<\infty$}/{FJ: persistent disagreement},
                      14.0/{$\kap\to\infty$}/{committed anchor}}{
  \draw[thick] (\x,-0.09) -- (\x,0.09);
  \node[below, font=\scriptsize] at (\x,-0.12) {\l};
  \node[below=3.5mm, font=\scriptsize\itshape, text=black!60] at (\x,-0.12) {\d};
}
\end{scope}
\end{tikzpicture}%
}
\caption{\textbf{One round through the belief layer.} A speaker's latent belief is rendered to text by the generator LLM; an appraiser LLM maps the text back to scalar evidence $e$; each listener takes one Bayesian update step modeled after the FJ update (Eq.~\ref{eq:fj}). $\kap$ places the agent on the classical stubbornness axis (bottom).}
\label{fig:schematic}
\end{figure*}

Large language models increasingly serve as the \emph{inhabitants} of social simulations: synthetic populations that discuss, persuade, and form collective opinions \citep{park2023generative,chuang2024simulating}. In most such systems the agent's opinion is revised by the LLM itself, in context. Two well-documented problems follow. First, populations tend to converge toward model-inherent biases despite assigned personas, a \emph{consensus collapse} that makes simulated societies unrealistically uniform and prompt-sensitive \citep{taubenfeld2024systematic,chuang2024simulating}. Second, a recent evidence-integration study finds that LLM confidence revisions violate Bayesian norms and are poorly calibrated under conflicting evidence \citep{kim2024evidence}.

Classical opinion dynamics offers the interpretable machinery these LLM-based simulations lack. DeGroot averaging \citep{degroot1974reaching} predicts consensus in connected populations; Friedkin--Johnsen (FJ) \citep{friedkin1990social} explains lasting disagreement through per-agent \emph{stubbornness}; committed-minority models study a small unyielding faction moving a flexible majority
\citep{xie2011social,centola2018experimental}. However, these models operate on bare scalars, with no notion of language which is precisely what LLMs now offer. We ask whether that language capability can be harnessed without losing the interpretability of the classical models.

\paragraph{Proposed architecture.} To combine the mathematical control of classical opinion dynamics with the language capabilities of LLM agents, we add a small, explicit belief layer between an agent's persona and its actions (i.e., speech generation for our set-up); we call the resulting agents \emph{Bayesian Chronicle Agents} (BCA) (Fig.~\ref{fig:schematic}). An agent's stance on a concept is a probability held outside the language model, the simplest instance of a concept--assertion \emph{chronicle}, our structured identity representation \citep{platnick2025idrag,alirezaie2024structural}. When the agent hears an utterance, a separate calibrated \emph{appraiser} model reads it into scalar evidence, and the belief takes one Bayesian update step (\S\ref{sec:layer}). A single per-agent knob, the prior strength $\kap$, sets how far each step moves the agent: a stubborn agent carries a large prior tally, so one observation barely moves it. With this architecture, \emph{what} an agent believes follows a transparent probabilistic rule; \emph{how} it acts/speaks in consequence is decided by the LLM.

The Bayesian belief update layer contributes the following to LLM-based social simulation. \textbf{(1) Recoverability.} We prescribe $\kap$, run the full language round-trip, and recover its rank order from the resulting belief dynamics (Spearman $1.0$ on every model tested; \S\ref{sec:kappa}). \textbf{(2) Regime fidelity.} One per-agent knob (at a global fixed $\gamma$, \S\ref{sec:layer}) sweeps the classical spectrum: DeGroot consensus ($\kap\to0$), FJ persistent disagreement (finite $\kap$), committed anchors ($\kap\to\infty$), each validated against the corresponding closed-form reference on four LLMs (\S\ref{sec:regimes}). \textbf{(3) Auditability.} Because every belief change is a logged event, we can compare what a speaker believed with what listeners were told, i.e., how faithfully the language channel transmits stance, surfacing a systematic distortion in every model we test and which regimes it corrupts. We release our code, prompts, and logged run data to support reproducibility.\footnote{\url{https://github.com/hafsa-akbar/bayesian-chronicle-agents}}
\section{Related Work}
\label{sec:related}

We situate our work at the intersection of three lines of research: 
\paragraph{Classical Opinion Dynamics.} DeGroot averaging \citep{degroot1974reaching}, Friedkin--Johnsen \citep{friedkin1990social}, and committed-minority models \citep{xie2011social,centola2018experimental}, introduced in \S\ref{sec:intro}, offer interpretable, mathematically tractable accounts of consensus, disagreement, and minority influence. A related line, bounded confidence \citep{hegselmann2002opinion}, instead lets agents ignore opinions too far from their own. These models are transparent by construction, but operate on bare scalars, with no notion of language which is exactly the gap an LLM agent can fill.

\paragraph{LLM-Based Opinion Simulation.} LLM agents converge toward model-inherent biases \citep{chuang2024simulating} and only partially toward assigned personas \citep{taubenfeld2024systematic}, yet can also reproduce classical signatures such as minority tipping \citep{ashery2025emergent}. At the individual level, LLM belief revision itself is found to violate Bayesian norms under conflicting evidence \citep{kim2024evidence}. Those dynamics \emph{emerge} from the model's prior; ours are prescribable per agent and, as we show, recoverable.

\paragraph{Explicit Belief State for LLM Agents.} Generative-agent memory architectures \citep{park2023generative} and structured identity models \citep{platnick2025idrag,alirezaie2024structural} persist \emph{in content} but leave belief revision to the LLM entirely or keep it static. The closest work to fill that gap is the concurrent Belief Engine \citep{yang2026belief}: an auditable log-odds accumulator with two empirical controls (evidence uptake, prior anchoring), evaluated through controlled single- and two-agent debates and human-trajectory replay. Our approach is complementary and differs on three axes: our single knob has an \emph{exact} classical correspondence rather than an empirical role; we test \emph{parameter recoverability} (prescribe, then recover through the language round-trip), unevaluated in Belief Engine; and we validate at \emph{population} scale against closed-form regime references.
  
\section{The Belief Layer}
\label{sec:layer}
The belief layer has three parts: a structured representation of what an agent believes (\S\ref{sec:identityrep}), a pair of LLM components that translate between belief and language (\S\ref{sec:belieftoaction}), and a Bayesian rule that updates belief from what the agent hears (\S\ref{sec:beliefupdate}).

\subsection{Identity Representation.} 
\label{sec:identityrep}
We represent agent identity as a graph of belief-holding \emph{concept} nodes (propositions the agent holds opinions about) held outside the language model, a simplified variation of the identity \emph{chronicle} i.e., knowledge graph learned from a person's digital footprint and shareable as a ``borrowable identity'' \citep{alirezaie2024structural}. Each concept carries an \emph{assertion set}, the competing natural-language stances on it (here an opposing pair $\{a^+, a^-\}$), and a \emph{credence}: a probability distribution over those assertions, whose dominant entry is the stance surfaced in agent behavior.

\subsection{From Belief to Action.}
\label{sec:belieftoaction}
Two LLM components connect the numeric belief state to natural language (Fig.~\ref{fig:schematic}). 
The \emph{generator} is the agent's voice: when the agent speaks, its current credence $b$ is rendered as a plain-language stance descriptor in the prompt (see App.~\ref{app:prompts} for the exact mapping), conditioning the generated action (here, the expressed opinion). The \emph{appraiser} is the agent's ears: when the agent hears another speaker, the appraiser reads the utterance and returns a single number $e\in[0,1]$, its judgment of how strongly the text supports $a^+$ over $a^-$ for the concept under discussion (the architecture supports appraising multiple concepts per utterance, but our single-concept experiments below exercise only one). Only this appraised evidence reaches the belief: the LLM controls what is \emph{said} and how it is \emph{understood}, while every change to what the agent \emph{believes} passes through the Bayesian update, keeping the dynamics controllable and auditable.

\subsection{The Belief Update} 
\label{sec:beliefupdate}

The layer has two parameters: prior strength $\kap=\alpha_0+\beta_0$ ($\alpha_0,\beta_0$ are the prior pseudo-counts the agent starts with), which encodes the agent's stubbornness, and a forgetting factor $\gamma\in(0,1]$ that sets how quickly old evidence fades relative to the founding prior. We hold $\gamma$ fixed at $0.7$ for every agent (chosen by a no-LLM sweep, App.~\ref{app:gamma}), so $\kap$ is the single per-agent knob we move; exact Bayes ($\gamma{=}1$) provably forces every population to consensus (App.~\ref{app:consensus}) and is kept as the no-forgetting ablation (\S\ref{sec:regimes}). Each heard utterance contributes one unit of evidence, split between the two stances by the appraised value $e$ and weighted by a fixed evidence weight $w$. We fix $w=1$ in all our experiments:

\begin{equation}
\begin{aligned}
\alpha &\leftarrow \alpha_0 + \gamma(\alpha-\alpha_0) + w e,\\
\beta  &\leftarrow \beta_0 + \gamma(\beta-\beta_0) + w(1-e).
\end{aligned}
\label{eq:update}
\end{equation}
Intuitively, the total pseudo-count $n=\alpha+\beta$ is how much the agent already ``knows''. Written in terms of the credence, the core of one step of Eq.~\ref{eq:update} is the convex blend
\begin{equation}
b' = (1-\eta)\,b + \eta\, e ,
\label{eq:fj}
\end{equation}
with susceptibility $\eta = w/n'$, where $n'$ is the post-discount count (App.~\ref{app:fjstep}): a stubborn agent (large $\kap$, hence large $n$) barely moves, a pliable one moves a lot. With forgetting on ($\gamma<1$), the exact step also adds a pull $\rho\,(b_0-b)$ back toward the agent's initial opinion---the defining ingredient to letting population dynamics like FJ persist. Closed forms and derivations are in App.~\ref{app:proofs}.
\section{Experimental Setup}
\label{sec:setup}

All runs use a single synthetic policy question with two opposing stances, so no model brings a strong real-world prior to it (concept, stances, and prompts in App.~\ref{app:prompts}). We run the experiment with $N{=}20$ agents on a complete graph, 5 seeds $\times$ 20 round-robin turns per condition (a fixed $\kappa$ assignment across agents, defined in \S~\ref{sec:results}). Each round every agent speaks once (generator temp.\ $0.9$); each comment is read once by the appraiser (temp.\ $0$), which turns it into the scalar evidence $e$ of Eq.~\ref{eq:update}, consumed by every listener. Finally, an independent per-round 0--100 self-report on that particular concept checks that agents faithfully express what they actually believe (probe prompt in App.~\ref{app:prompts}; audit in App.~\ref{app:channel}).

We repeated our experiments with \texttt{gpt-5.4-mini}, \texttt{gpt-5.4} (OpenAI), \texttt{Llama-4-Scout-17B-16E} (open weights, Meta), and \texttt{claude-sonnet-4-6} (Anthropic), under the same sampling parameters across all four models. We found LLM appraisers tend to state overconfident probabilities, so we fit one temperature $\tau$ per model on 100 labelled utterances, reducing held-out calibration error (App.~\ref{app:calibration}).
\section{Results}
\label{sec:results}
\begin{figure*}[t]
\centering
\includegraphics[width=\textwidth]{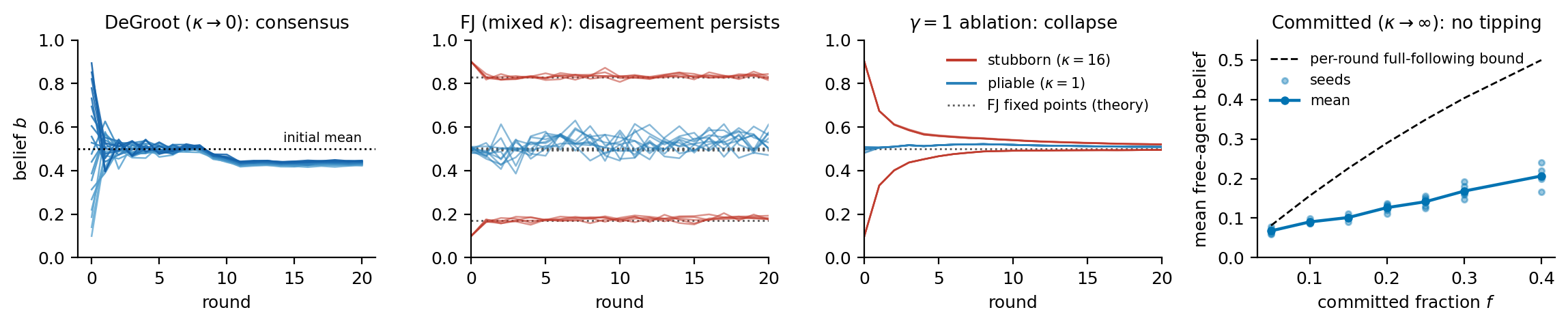}
\caption{\textbf{One knob, three classical regimes}
(\texttt{gpt-5.4-mini}; one line per agent, averaged over seeds; all
four models replicate these regimes, Tab.~\ref{tab:crossmodel}).
\textbf{(a)}~When everyone is pliable, opinions merge into a single
consensus (the late drift is mild channel bias,
App.~\ref{app:channel}). \textbf{(b)}~Add stubborn agents and
disagreement persists, settling onto the dotted theory-predicted
levels. \textbf{(c)}~Turn forgetting off and the same population
collapses to consensus anyway. \textbf{(d)}~A minority that never
budges pulls the majority smoothly, with no tipping point.}
\label{fig:regimes}
\end{figure*}

\subsection{Prescribed $\kap$ is recoverable through the language channel}
\label{sec:kappa}
Is the prescribed stubbornness recoverable from behaviour after the full language round-trip? For each model we sweep
$\kap\in\{0.5,1,2,4,8,16,32\}$ (14{,}000 utterances, 266{,}000 listener events per model) and invert the exact one-step identity (App.~\ref{app:proofs}) per event: a listener at belief $b$ who hears a speaker with latent belief $s$ and moves to $b'$ yields
\begin{equation}
\hat\kap = \frac{w\,(s-b) - (b'-b)\,(\gamma m + w)}{(b'-b) - (1-\gamma)\,(b_0-b)},
\label{eq:inversion}
\end{equation}

with $m$ the listener's discounted evidence count and $b_0$ its initial belief (notation, derivation, and numerical-exactness check in App.~\ref{app:proofs}). Using the speaker's latent belief $s$ instead of the appraised evidence $e$ (which would recover $\kap$ by construction) means recovery succeeds only if the generated text carried that belief and the appraiser read it back out, so any remaining error is attributable to the language channel. We discard ill-conditioned events where the listener already nearly agrees with the speaker ($|s-b| \leq 0.05$): such utterances barely move the listener regardless of $\kap$, so they carry no information about stubbornness (retention statistics in App.~\ref{app:gamma}).

On all four models the channel is faithful (appraised evidence tracks latent speaker belief, Pearson $0.94$--$0.97$) and $\kap$ is recovered \emph{perfectly} in rank order (Spearman 1.0 on every model; Fig.~\ref{fig:kappa} and per-model numbers in App.~\ref{app:channel}). Recovered magnitudes are uniformly attenuated (\eg $\kap{=}32 \to 25$--$30$; per-condition medians in Tab.~\ref{tab:kappaperk}): the channel inflates the evidence--belief gap (App.~\ref{app:channel}), so agents appear somewhat more pliable than prescribed. Thus, within BCA, prescribed $\kap$ remains behaviorally recoverable after passing through the language channel.

\subsection{One knob, three classical regimes}
\label{sec:regimes}
We dial only $\kap$ and ask whether the population produces each canonical regime, judged against the corresponding closed-form reference (Fig.~\ref{fig:regimes}; full cross-model numbers in Tab.~\ref{tab:crossmodel}, App.~\ref{app:channel}). 

\textbf{Consensus (pliable limit $\kap\to0$; here $\kap{=}0.1$).} When no one is stubborn, DeGroot theory predicts consensus near the average initial opinion. Spread indeed collapses by three orders of magnitude on all four models, but \emph{where} consensus lands exposes the channel: each channel misreads stances in a characteristic direction, and pliable agents---believing whatever they hear---accumulate the misreading. Against the true initial mean of $0.50$, consensus lands at $0.50$ (\texttt{gpt-5.4}) and $0.44$ (\texttt{gpt-5.4-mini}), whose symmetric misreadings largely cancel, but at $0.15$ (\texttt{claude-sonnet-4-6}) and $0.01$ (\texttt{Llama-4-Scout}), whose one-sided misreadings compound. An end-to-end simulation would report these displaced consensuses as findings; the belief layer instead detects and measures the bias (App.~\ref{app:channel}).

\textbf{Persistent disagreement (FJ; stubborn camps $\kap{=}16$ at the extremes, pliable $\kap{=}1$ agents between).} Once some agents are stubborn, opinions should stop merging: camps hold their ground and pliable agents settle between them. We observe this stable spread, with per-agent final beliefs matching the FJ fixed point at $R^2 = 0.93$--$0.99$ across models. At $\gamma{=}1$ the same population collapses back to near-consensus and $\kap$-recovery degrades sharply, showing forgetting is the enabling ingredient (\S\ref{sec:layer}; full ablation on \texttt{gpt-5.4-mini}: App.~\ref{app:gamma}, Tab.~\ref{tab:gamma}).

\textbf{Minority influence (committed $\kap\to\infty$ minority vs.\ free $\kap{=}4$ majority).} The affine update provably cannot produce a tipping point (App.~\ref{app:notipping}), and none appears: sweeping the committed fraction from $5\%$ to $40\%$, the majority's mean rises smoothly on every model (Fig.~\ref{fig:regimes}d), in contrast to critical-mass experiments \citep{xie2011social,centola2018experimental}.
\section{Conclusion}
A Bayesian belief layer makes LLM social simulation controllable in the currency the opinion-dynamics field already trusts: prescribed $\kap$ remains recoverable through the full language round-trip in exact rank order (Spearman $1.0$ on four models), it \emph{controls} the classical regime spectrum against closed-form references, and its logged events make simulation \emph{auditable}, surfacing per-model channel bias. Multi-concept identities, directional (not just confidence) channel correction, and heterogeneous topologies are natural next steps.

\section*{Limitations}
Our study isolates the mechanism at the cost of scope: one synthetic topic, a complete graph, $N{=}20$ agents, binary stances, and validation against classical theory rather than human trajectories, so these simulations are not predictions of human opinion change. Our engine is also sequential (one utterance at a time), whereas the theoretical reference is synchronous mean-field; nevertheless, terminal beliefs match the corresponding FJ fixed points at $R^2=0.93\text{--}0.99$ across models, suggesting that this mismatch is small in the setting studied (App.~\ref{app:consensus}). The appraiser is itself an LLM: calibration corrects the appraiser's confidence, but directional distortion of the channel (whether introduced in rendering or reading) passes through and still shapes dynamics in the fully pliable regime. On the data side, all conversation in our runs is model-generated by construction; the only human-labelled data is the small, single-topic calibration set of App.~\ref{app:calibration}, so the appraiser's calibration is only as strong as those 100 judgments. And because the topic is deliberately fictional, chosen so that models bring no strong pre-trained stance, our results do not speak to debates where they do. Finally, our pipeline requires two LLM calls per utterance across four models, which may limit exact reproducibility without comparable API access.

\bibliography{custom}
\newpage
\appendix

\section{Choosing $\gamma$: the oracle sweep and full ablation}
\label{app:gamma}

\begin{figure}[h!]
\centering
\includegraphics[width=0.9\columnwidth]{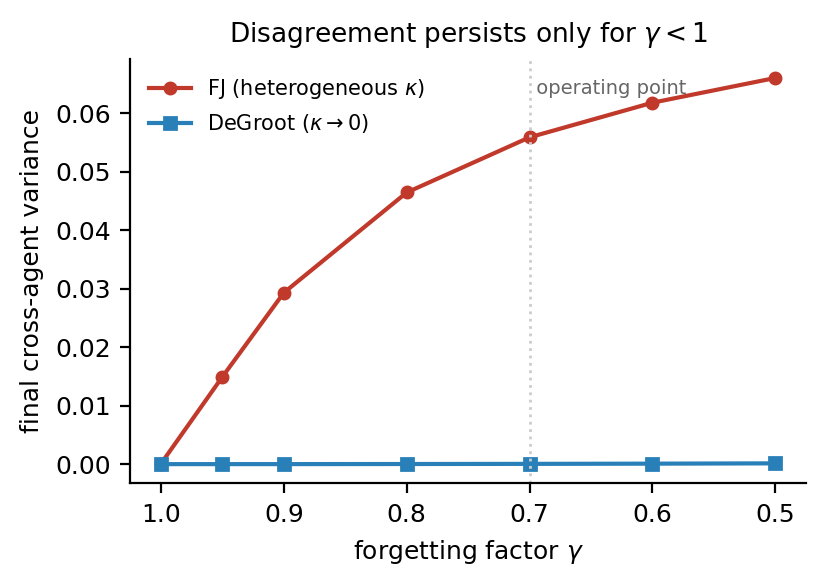}
\caption{Oracle $\gamma$ sweep (belief mechanism only, no LLM calls):
final cross-agent variance in the FJ configuration rises as soon as
$\gamma<1$ and saturates near the chosen operating point, while the
DeGroot configuration reaches consensus for every $\gamma$.}
\label{fig:gammasweep}
\end{figure}

A mechanism-only sweep (oracle evidence, no LLM calls) over $\gamma\in\{1.0,0.95,0.9,0.8,0.7,0.6,0.5\}$ shows final FJ cross-agent variance rising from $9{\times}10^{-5}$ at $\gamma{=}1$ to $0.056$ at the $\gamma{=}0.7$ operating point ($0.066$ by $\gamma{=}0.5$), while DeGroot remains at consensus for every $\gamma$ (Fig.~\ref{fig:gammasweep}). We fix $\gamma{=}0.7$ \emph{before} any paid LLM run and hold it constant across all models and regimes. Table~\ref{tab:gamma} reports the full ablation through the language channel.

\begin{table}[h!]
\centering\small
\begin{tabular}{lcc}
\toprule
Metric (language channel) & $\gamma{=}0.7$ & $\gamma{=}1$ \\
\midrule
FJ final variance / persists & $\mathbf{0.053}$ / yes & $10^{-4}$ / no \\
FJ $R^2$ vs.\ fixed point & $\mathbf{0.97}$ & $-1.1$ \\
$\kap$-recovery rel.\ error & $\mathbf{0.49}$ & $1.41$ \\
Well-conditioned events & $\mathbf{0.76}$ & $0.12$ \\
DeGroot final variance & $5{\times}10^{-5}$ & $7{\times}10^{-7}$ \\
Belief--slider audit ($r$) & $0.99$ & $0.98$ \\
\bottomrule
\end{tabular}
\caption{Forgetting ablation (\texttt{gpt-5.4-mini}, 5 seeds). Removing forgetting kills FJ persistence and substantially degrades $\kap$-recoverability; consensus and auditability are unaffected}
\label{tab:gamma}
\end{table}

Retention of well-conditioned events after the \S\ref{sec:kappa} filters rises monotonically with prescribed stubbornness (\texttt{gpt-5.4-mini}: $0.45$ at $\kap{=}0.5$ to $0.92$ at $\kap{=}32$, closely matched by the other three models): low-$\kap$ conditions are noisier, not selectively discarded.

\section{Appraiser calibration per model}
\label{app:calibration}
The 100 utterances were LLM-generated to span all stance bands, matching the medium the appraiser reads in deployment. Each calibration utterance carries a \emph{soft label}: a graded gold target $y\in[0,1]$ for its stance strength, assigned by hand by a single human annotator (\eg $0.8$ for a clear but not absolute lean toward $a^+$, $0.5$ for balanced); each label carries a one-line rationale in the released data.

The appraiser states $p_+$; calibration applies temperature scaling \citep{guo2017calibration}, $p_{\mathrm{cal}} = \sigma(\mathrm{logit}(p_+)/\tau)$, with one $\tau$ per model fit by minimising the cross-entropy $-[y\log p + (1-y)\log(1-p)]$ on an 80-example split and evaluated on 20 held-out utterances.

Validation ECE improves $0.101\to0.052$ (\texttt{gpt-5.4-mini}), $0.103\to0.046$ (\texttt{gpt-5.4}), $0.040\to0.034$ (\texttt{Llama-4-Scout}), $0.081\to0.033$ (\texttt{claude-sonnet-4-6}). Calibration is independent of $\kap$ and $\gamma$; each fit is frozen and reused by all downstream runs of that model.

\section{Auditing the language channel}
\label{app:channel}

\begin{figure*}[t!]
\centering
\includegraphics[width=\textwidth]{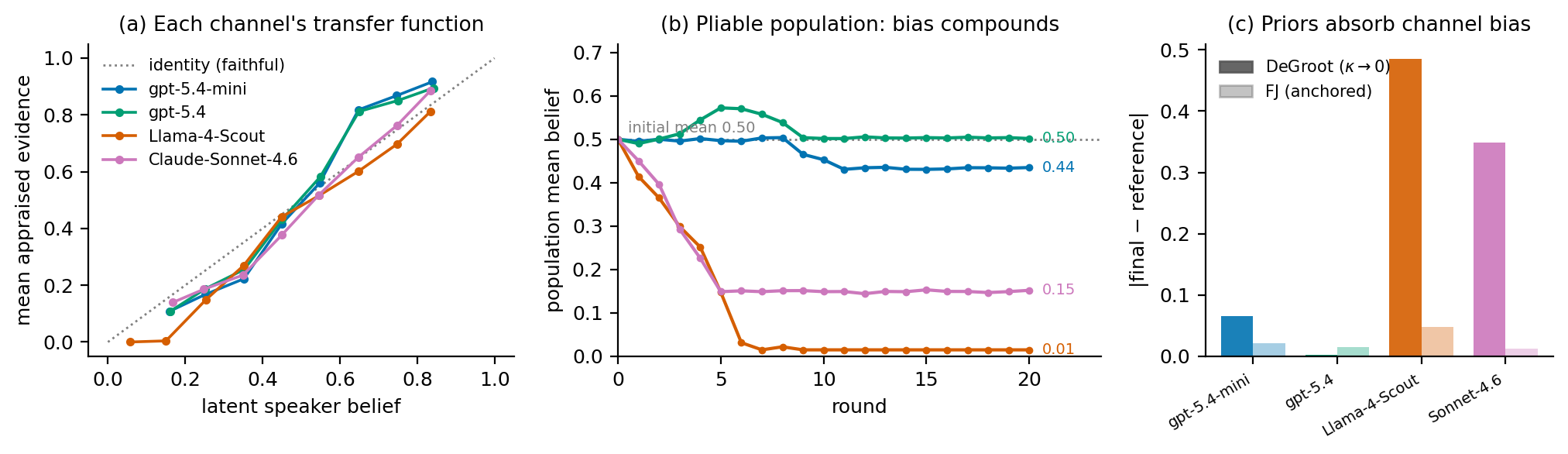}
\caption{\textbf{The belief layer as a measurement instrument.}
(a)~Every model's channel transfer function deviates from identity, each in its own way (GPTs: symmetric expansion; Llama: uniform downward shift; Claude: asymmetric negative-side amplification).
(b)~Pliable DeGroot populations compound asymmetric distortion into a
confidently wrong consensus (final values labelled; reference $0.50$).
(c)~Displacement from the theoretical reference: large for the pliable
regime, small for the prior-anchored FJ regime on the same
models---stubbornness absorbs channel bias.}
\label{fig:channel}
\end{figure*}

\paragraph{How each model transmits stances.} For every utterance the layer logs two numbers: the belief the speaker actually held ($s$) and the evidence the appraiser reported to listeners ($e$). Averaging $e$ against $s$ draws each model's \emph{channel transfer curve} (Fig.~\ref{fig:channel}a); a perfectly faithful channel would sit on the diagonal. All models show some deviation. The GPT channels \emph{exaggerate}: moderate stances arrive as more extreme, on both sides of neutral. \texttt{Llama-4-Scout} transmits every stance as leaning somewhat more toward $a^-$, its whole curve sitting below the diagonal. The Claude channel exaggerates only the $a^-$ side (\eg a mild $0.35$ stance arrives as $0.24$) while staying roughly faithful on the $a^+$ side. These are properties of how each model \emph{transmits} stance through the generate--appraise round-trip, not opinions about the topic itself. The curve measures the generator and appraiser \emph{jointly}; separating rendering from reading (\eg by cross-appraising one model's utterances with another model's appraiser) is left to future work.

\paragraph{Why the shape of the misreading matters.} A symmetric exaggeration pushes some readings up and others down, so over a balanced population the errors cancel and the consensus stays put---which is why the GPT populations land near $0.50$ (Fig.~\ref{fig:channel}b). A shifted or one-sided curve instead injects a small push in the \emph{same} direction every round. Pliable agents have nothing to resist it with, so the pushes accumulate into the large drifts of Llama and Claude; stubborn agents are re-anchored by their priors at every step, so the same push cannot accumulate, and the FJ regime lands near theory on all four models (Fig.~\ref{fig:channel}c).

\paragraph{Why appraiser calibration addresses a different failure mode.} Temperature scaling fixes \emph{overconfidence}: an appraiser that tags evidence $e=0.9$ on an utterance that supports the assertion only at $0.7$ is pulled back toward the label. But it rescales confidence symmetrically about the neutral point, so it cannot raise, lower, or bend one side of the transfer curve: \emph{directional} bias of the language channel passes through. That bias is an artefact of the channel, not of the belief-update layer, and, because every event is logged, it is measurable in our architecture rather than silently absorbed.

\paragraph{Does the hidden belief govern behaviour?} A separate check: each round every agent also gives an independent 0--100 self-report of its stance (App.~\ref{app:prompts}), which never enters any belief update. These self-reports track the hidden belief of an agent closely ($r\approx0.98$--$0.99$ over $18{,}000$ reports per model), so the Beta state is not internal bookkeeping: it is what the agent expresses in a persona-coherent manner which is verified through a channel separate from the generator--appraiser loop.

Table~\ref{tab:crossmodel} collects the headline metrics for all four models.

\begin{table*}[t!]
\centering\small
\setlength{\tabcolsep}{4pt}
\begin{tabular}{lcccc}
\toprule
& \texttt{gpt-5.4-mini} & \texttt{gpt-5.4} & \texttt{Llama-4-Scout}
& \texttt{claude-sonnet-4-6} \\
\midrule
Weights / provider & closed / OpenAI & closed / OpenAI & open / Meta
& closed / Anthropic \\
Appraiser temperature $\tau$ & 1.89 & 2.15 & 1.51 & 1.69 \\
Channel alignment (Pearson) & 0.96 & 0.97 & 0.94 & 0.96 \\
DeGroot consensus (ref.\ 0.50) & 0.44 & 0.50 & 0.01 & 0.15 \\
FJ persists / $R^2$ & yes / 0.97 & yes / 0.98 & yes / 0.93
& yes / 0.99 \\
Committed: smooth, no tipping & yes & yes & yes & yes \\
$\kap$-recovery (Spearman) & 1.00 & 1.00 & 1.00 & 1.00 \\
$\kap$-recovery (rel.\ error) & 0.49 & 0.46 & 0.33 & 0.27 \\
Auditability, belief--slider ($r$) & 0.99 & 0.99 & 0.98 & 0.99 \\
\bottomrule
\end{tabular}
\caption{Full pipeline across four models ($\gamma{=}0.7$, 5 seeds, 20
rounds, per-model $\tau$ refit). Structural results replicate on all
four; $\kap$ rank-recovery is Spearman $1.0$ for every individual
seed, not only the pooled per-condition medians of
Fig.~\ref{fig:kappa}.}
\label{tab:crossmodel}
\end{table*}

\begin{table}[t]
\centering\small
\setlength{\tabcolsep}{4.5pt}
\begin{tabular}{lrrrrrrr}
\toprule
prescribed $\kap$ & 0.5 & 1 & 2 & 4 & 8 & 16 & 32 \\
\midrule
\texttt{gpt-5.4-mini}      & 0.0 & 0.3 & 1.0 & 2.5 & 5.6 & 12.1 & 25.3 \\
\texttt{gpt-5.4}           & 0.0 & 0.3 & 1.0 & 2.6 & 6.0 & 12.6 & 26.2 \\
\texttt{Llama-4-Scout}     & $-0.1$ & 0.4 & 1.6 & 3.8 & 7.6 & 14.8 & 30.0 \\
\texttt{claude-sonnet-4-6} & 0.3 & 0.5 & 1.3 & 3.4 & 7.0 & 14.0 & 28.7 \\
\bottomrule
\end{tabular}
\caption{Recovered stubbornness (oracle-aligned; per-condition medians pooled over seeds) for every prescribed $\kap$. Rank order is exact in every row; magnitudes are uniformly attenuated, increasingly so at low $\kap$.}
\label{tab:kappaperk}
\end{table}

\begin{figure}[t]
\centering
\includegraphics[width=0.8\columnwidth]{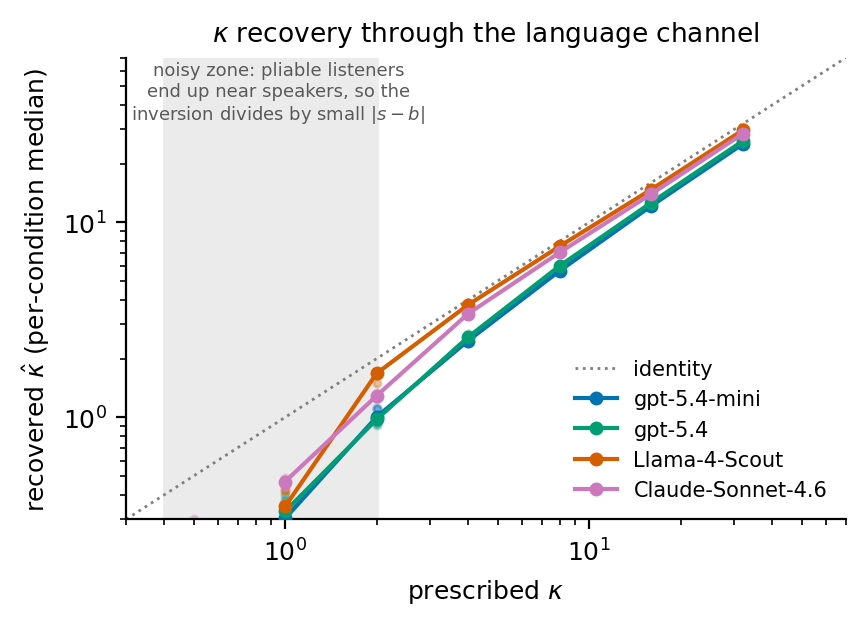}
\caption{\textbf{Prescribe-then-recover: $\kap$ is recoverable through the language channel.} Prescribed stubbornness (x) vs.\ value recovered from listeners' belief movements (Eq.~\ref{eq:inversion}). Rank order is exact on all four
models; magnitudes are uniformly attenuated by the channel
(Tab.~\ref{tab:kappaperk}).}
\label{fig:kappa}
\end{figure}

\section{Prompts and protocol}
\label{app:prompts}

\textbf{Concept.} The simulated debate concerns one municipal policy
question for the fictional town of Aldenvale
(\texttt{transit\_priority}), with the opposing assertion pair
$a^+$: ``expand rail'' and $a^-$: ``keep roads''. The town and the
question are invented so that no model brings a strong pre-trained
prior to either side.

\textbf{Generator} (system): ``You are a resident of the fictional town of Aldenvale taking part in a community discussion about local transportation policy. You speak naturally, like a real person at a town
meeting.'' (user): the two stances, a natural-language descriptor of the agent's private stance strength, and an instruction to write a 1--2 sentence comment without mentioning numbers, probabilities, or internal variables.
\textbf{Appraiser} (system): ``You are an impartial stance classifier
\ldots judging the evidence in the text rather than your own opinion.'' 
(user): both stances, the statement, and a request for \texttt{\{"p\_plus": <0..1>\}}. \textbf{Probe}: an independent request for a single 0--100 integer stance report. Full templates, seeds, and the round-robin event loop are in the released code.

The stance descriptor is a fixed nine-bin qualitative mapping from $b$ to phrases, from ``completely opposed to rail expansion and fully committed to roads'' ($b<0.10$) through ``genuinely torn and balanced'' ($0.45<b\le0.55$) to ``completely committed to rail expansion'' ($b\ge0.90$); no numbers appear in any prompt.

\section{Derivations and proof sketches}
\label{app:proofs}

\paragraph{Notation.} An agent's state on the concept is the pair of
pseudo-counts $(\alpha,\beta)$: accumulated evidence for $a^+$ and
$a^-$ respectively. The total count is $n=\alpha+\beta$ and the
credence is $b=\alpha/n$. The prior counts are $(\alpha_0,\beta_0)$
with prior strength $\kap=\alpha_0+\beta_0$ and prior mean
$b_0=\alpha_0/\kap$ (the agent's initial stance). Each heard utterance
delivers appraised evidence $e\in[0,1]$ with weight $w$; $c$ counts how
many utterances the agent has absorbed; $\gamma\in(0,1]$ is the
forgetting factor. Subscripts index update steps: $n_c$ is the count
after $c$ updates.

\subsection{The count is deterministic}
\label{app:count} 
Summing the two lines of Eq.~\ref{eq:update} makes the evidence $e$ cancel:
\begin{equation*}
n_{c+1} = \kap + \gamma\,(n_c-\kap) + w .
\end{equation*}
Unrolling from $n_0=\kap$ gives
\begin{equation*}
n_c =
\begin{cases}
\kap + w\,\tfrac{1-\gamma^{c}}{1-\gamma} & \gamma<1,\\[2pt]
\kap + wc & \gamma=1 .
\end{cases}
\end{equation*}
So the count, that is how much the agent ``already knows'', is a known
function of $(\kap,\gamma,c)$, independent of what was said. This is
the fact the $\kap$-recovery estimator exploits (\S\ref{app:inversion}). For $\gamma<1$
the count saturates at $n^\star=\kap+w/(1-\gamma)$; for $\gamma=1$ it grows forever.

\subsection{One step is a Friedkin--Johnsen update}
\label{app:fjstep}
Substituting $\alpha = b\,n$ into the $\alpha$-line of Eq.~\ref{eq:update} and
dividing by $n_{c+1}$ yields the exact per-step identity
\begin{equation*}
b' \;=\; b \;+\; \eta_c\,(e-b) \;+\; \rho_c\,(b_0-b),
\end{equation*}
with susceptibility $\eta_c = w/n_{c+1}$ and anchor weight $\rho_c = \kap(1-\gamma)/n_{c+1}$. Each utterance therefore moves the credence toward the evidence, plus (when forgetting is on) a pull back toward the agent's \emph{initial} stance---exactly the two defining ingredients of the FJ update: susceptibility to the social signal and anchoring to one's founding opinion \citep{friedkin1990social}. Stubbornness enters as claimed: $\eta_c$ is strictly decreasing in $\kap$. At $\gamma{=}1$ the anchor vanishes ($\rho_c=0$) and the step reduces exactly to the two-term blend of Eq.~\ref{eq:fj} with $\eta = w/(n_c+w)$.

\subsection{Exact Bayes forces consensus; forgetting prevents it}
\label{app:consensus}
At $\gamma{=}1$ the update accumulates evidence without discount, so
after $t$ updates
\begin{equation*}
b^{(t)} = \rho^{(t)} b_0 + \bigl(1-\rho^{(t)}\bigr)\,\bar e^{(t)},
\end{equation*}
where $\rho^{(t)} = \kap/(\kap+wt)$ is the residual prior weight and
$\bar e^{(t)}$ the running average of the evidence heard. The
prior weight $\rho^{(t)}\to 0$: every agent's stance converges onto the
shared evidence stream, so on a connected graph disagreement is
transient---no assignment of $\kap$ can sustain it. For $\gamma<1$ the count saturates (\S\ref{app:count}), so $\eta$ stays bounded away from zero and setting $b'=b$ in the identity of \S\ref{app:fjstep} gives the stationary point
\begin{equation*}
b^\star = \lambda_0\, b_0 + (1-\lambda_0)\,\bar e, \quad
\lambda_0 = \tfrac{\kap(1-\gamma)}{\kap(1-\gamma)+w},
\end{equation*}
the classical FJ steady state: the initial stance keeps a
\emph{permanent} weight that grows with stubbornness, so heterogeneous
$\kap$ sustains heterogeneous opinions forever. ($\lambda_0$ is the
equilibrium mixture weight $\rho_\infty/(\rho_\infty+\eta_\infty)$,
not the limit of the per-step $\rho_c$ of \S\ref{app:fjstep}.)

For the population reference of Fig.~\ref{fig:regimes}b and the $R^2$
of \S\ref{sec:regimes}: with $N$ agents on the complete graph,
zero-diagonal uniform weights $W_{ij}=1/(N{-}1)$, and per-agent
susceptibility $\lambda_i = w/(\kap_i(1-\gamma)+w)$, the theory lines
are the \emph{synchronous} (mean-field) FJ fixed point
$\mathbf b^\star = (I-\Lambda W)^{-1}(I-\Lambda)\,\mathbf b_0$ with
$\Lambda=\mathrm{diag}(\lambda_i)$, i.e.\ the stationary belief vector
if every agent updated from one simultaneous snapshot of its
neighbours, rather than our engine's sequential, one-utterance-at-a-time
schedule. $R^2$ compares per-agent terminal
beliefs (5-seed averages) against $\mathbf b^\star$ directly, without
refitting ($1-\mathrm{SS}_{\mathrm{res}}/\mathrm{SS}_{\mathrm{tot}}$,
no intercept); that the fit is $0.93$--$0.99$ despite this mismatch
indicates the sequential dynamics track the mean-field reference
closely.

\subsection{Recovering $\kap$ from behavior}
\label{app:inversion}
This derives Eq.~\ref{eq:inversion}. Take the exact one-step identity of \S\ref{app:fjstep}, $b' = b + (w/n')(e-b) + (\kap(1-\gamma)/n')(b_0-b)$ with $n' = \kap + \gamma m + w$, where $m = w(1-\gamma^{c})/(1-\gamma)$ is known from the listener's event history (\S\ref{app:count}) and $b_0$ is its logged initial belief. Substituting the speaker's latent belief $s$ for the evidence $e$ (the channel test of \S\ref{sec:kappa}) and solving the resulting linear equation for $\kap$ gives Eq.~\ref{eq:inversion}. Because the identity is exact at every $\gamma$, so is the inversion: applied to the logged events with the consumed evidence $e$ it returns the prescribed $\kap$ to numerical precision on all four models, and at $\gamma{=}1$ it reduces to the familiar $\hat\kap = w\hat\eta^{-1} - m - w$ with $\hat\eta=(b'-b)/(s-b)$. All deviation that remains when $s$ replaces $e$ is therefore attributable to the language channel.

\subsection{No tipping point}
\label{app:notipping}
Consider the complete graph with a fraction $f$ of agents committed (frozen) at $b{=}1$ and the free agents sharing $\kap$ and a common starting stance $b_F^{(0)}$. By symmetry all free agents hold a common belief $b_F^{(t)}$, and the average evidence a free agent hears in round $t$ is
$f\cdot 1 + (1-f)\,b_F^{(t)}$ (exactly $fN/(N{-}1)$ once a free agent's
own belief is excluded from its neighbours; with $N{=}20$ the two are
indistinguishable). Substituting into the blend gives
\begin{align*}
1-b_F^{(t+1)} &= \bigl(1-\eta_t f\bigr)\bigl(1-b_F^{(t)}\bigr),\\
b_F^{(T)} &= 1 - \bigl(1-b_F^{(0)}\bigr)\!\prod_{t<T}\!\bigl(1-\eta_t f\bigr).
\end{align*}
As a function of $f$ this is continuous, smooth, and strictly increasing: no discontinuity, no bistability, no critical mass. (With forgetting on, the one-step coefficients gain the anchor term $\rho_t(b_F^{(0)}-b_F^{(t)})$ of \S\ref{app:fjstep} and remain affine in $f$; composing $T$ such steps makes $b_F^{(T)}$ polynomial, hence still smooth, in $f$.) Tipping requires a nonlinearity (\eg acceptance thresholds or majority rules) that an affine update does not contain; the experiments confirm the predicted smooth dose--response. The dashed curve in Fig.~\ref{fig:regimes}d is a \emph{stylised full-following reference}, not a proven bound: it instantiates one aggregated unit-weight update per round, an idealisation of the sequential engine (which applies $N{-}1$ per-utterance updates per round); with forgetting on, the anchor term of \S\ref{app:fjstep} additionally pulls free agents toward their starting stance. Both effects place the measured dose--response below this reference curve in our runs.

\end{document}